\documentclass[twocolumn]{aastex63}
\usepackage{epsfig}
\usepackage{natbib}
\usepackage{amsmath}
\usepackage{hyperref}
\usepackage{lineno}

\newcommand{\kms}{\mbox{${\rm km~s^{-1}}$}}
\newcommand{\kmsMpc}{\mbox{km s$^{-1}$ Mpc$^{-1}$}}

\begin{document}

\title{Ho'oleilana: An Individual Baryon Acoustic Oscillation?}

\author{R. Brent Tully}
\affil{Institute for Astronomy, University of Hawaii, 2680 Woodlawn Drive, Honolulu, HI 96822, USA}
\author{Cullan Howlett}
\affil{School of Mathematics and Physics, The University of Queensland, Brisbane, QLD 4072, Australia.}
\author{Daniel Pomar\`ede}
\affil{Institut de Recherche sur les Lois Fondamentales de l'Univers, CEA Universit\'e Paris-Saclay, 91191 Gif-sur-Yvette, France}

\begin{abstract}
Theory of the physics of the early hot universe leads to a prediction of baryon acoustic oscillations that has received confirmation from the pair-wise separations of galaxies in samples of hundreds of thousands of objects. Evidence is presented here for the discovery of a remarkably strong {\it individual} contribution to the baryon acoustic oscillation (BAO) signal at $z=0.068$, an entity that is given the name Ho'oleilana.  The radius of the 3D structure is $155\,h_{75}^{-1}$~Mpc.  At its core is the Bo\"otes supercluster. The Sloan Great Wall, CfA Great Wall, and Hercules complex all lie within the BAO shell.  
The interpretation of Ho'oleilana as a BAO structure with our preferred analysis implies a value of the Hubble constant of $76.9^{+8.2}_{-4.8}\,\kmsMpc.$

\end{abstract}

\section{Introduction}

Pressure waves generated in the hot plasma of the early universe become imprinted in baryon fluctuations approximately 390,000 years after the hot Big Bang \citep{Peebles+70, Sunyaev+70}.  The remnants of these waves create a ruler that, observed across time in the evolving universe, provides constraints on the physics governing cosmic evolution \citep{Weinberg+13,Aubourg+15}.  
\citet{Eisenstein+98} investigated the possibility that early universe fluctuations caused by the baryon component of matter might explain structure on scales of $\sim13,000$~\kms\ \citep{Tully86,Tully+92,Broadhurst+90} and hints of baryon induced features in the power spectrum of galaxy correlations were first announced by \citet{Percival+01}.
Subsequently, compelling evidence for what have come to be called baryon acoustic oscillations (BAO) has been seen as a peak in the pair-wise separations of galaxies throughout cosmic history \citep{Cole+05,Eisenstein+05,Beutler+11,Blake+11,Ross+15,Alam+17,Alam+21}.  In all published cases, the BAO feature is a $statistical$ imprint compounded by contributions from many locations. 

Studies such as \cite{Scrimgeour2012} and \cite{Goncalves2018} have identified the scale at which the Universe reaches one percent homogeneity as $\sim 70-120\,h_{75}\,\mathrm{Mpc}$. By logical arguments, the density fluctuations anticipated in individual BAO shells (which exist on scales larger than the homogeneity scale) can then be only a few percent of the mean matter density. So it has not been expected that individual BAO can be discerned. It was demonstrated by \citet{arnalte-mur+12}, though, that assuming BAO developed out of pre-recombination central dark matter concentrations identifiable today as rich clusters, the scales of associated BAO could be identified by wavelet analysis and the stacked density maps from $\sim 800$ centers. These centers can be further studied to identify the structures that contribute most substantially to the total BAO signal.

\begin{figure}[!]
\centering
\includegraphics[width=.97\linewidth]
{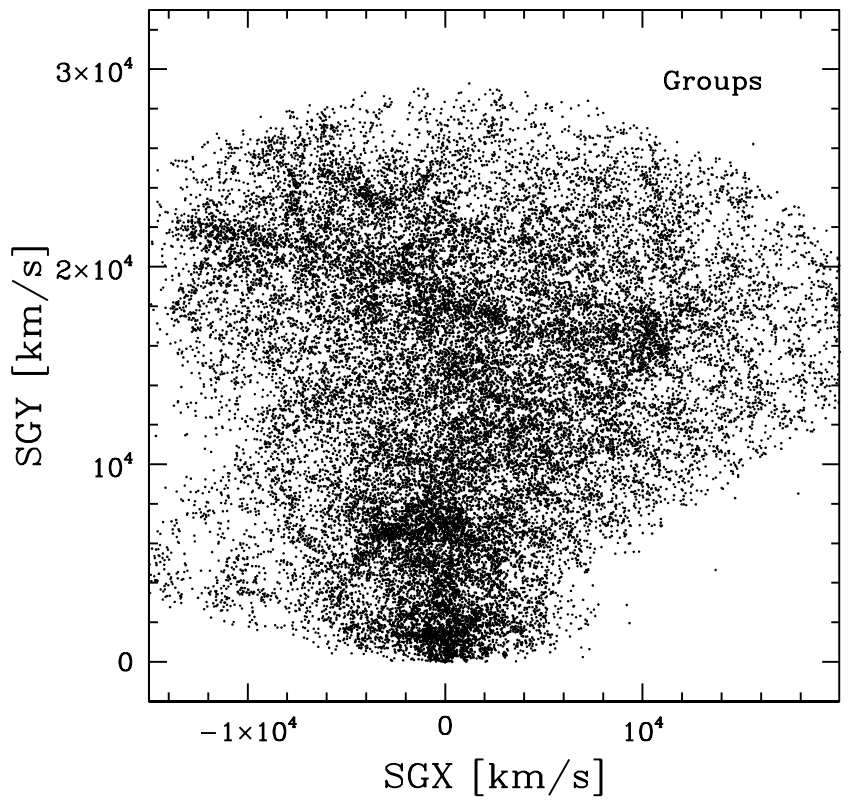}
\includegraphics[width=.97\linewidth]
{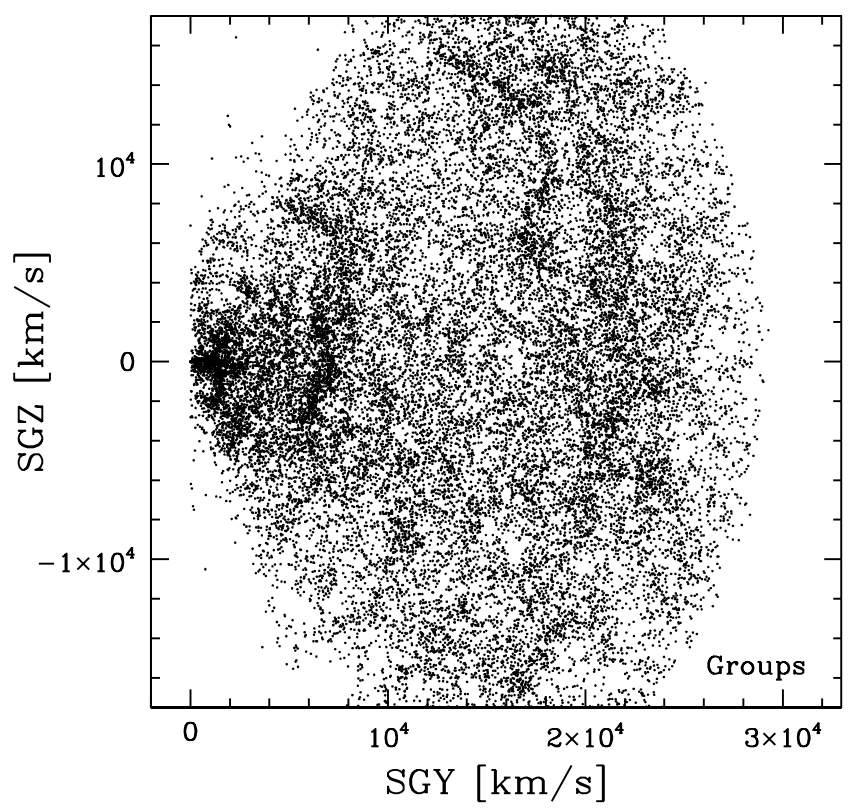}
\caption{
Zoom of Fig 21 from \citet{Tully+23} in the galactic north sector showing the distribution of Cosmicflows-4 galaxy groups in supergalactic coordinates. 
}
\label{fig:2views}
\end{figure}

We were not looking for BAO.  However visual examination of maps from the Cosmicflows-4 compilation of galaxy distances \citep{Tully+23} revealed a structure that invited further inspection.  By way of introduction, the two orthogonal views in supergalactic coordinates in Figure~\ref{fig:2views} show the distribution of galaxy groups north of the Milky Way equator in this data set.\footnote{Distances are given in units of CMB frame velocities, $V_{cmb}$.  Distances in Mpc, $d$, are directly related: $d=f(\Omega_m,\Omega_{\Lambda}) V_{cmb}/H_0$ where $f(\Omega_m,\Omega_{\Lambda})$ is a small adjustment dependent on cosmological model.}
The SGY axis roughly tracks redshifts. An evident overdensity is seen at SGY$\sim 20,000$~\kms, part of which is the Sloan Great Wall \citep{Gott+05}. The Center for Astrophysics Great Wall \citep{deLapparent+86} is seen at SGY$\sim 7000$~\kms.

\begin{figure}[!]
\centering
\includegraphics[width=.97\linewidth]{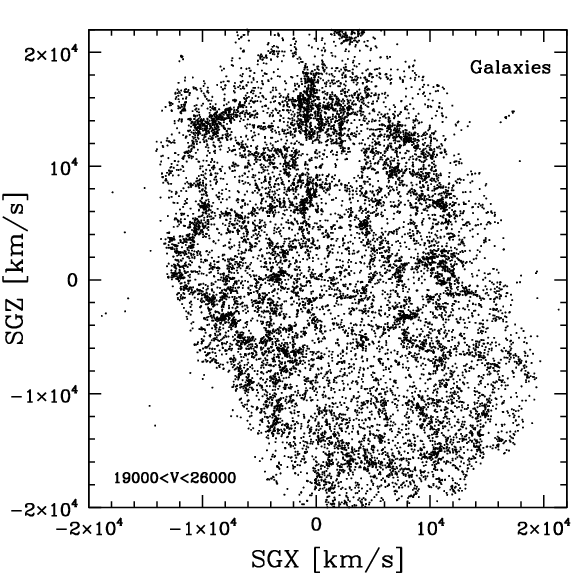}
\includegraphics[width=.97\linewidth]{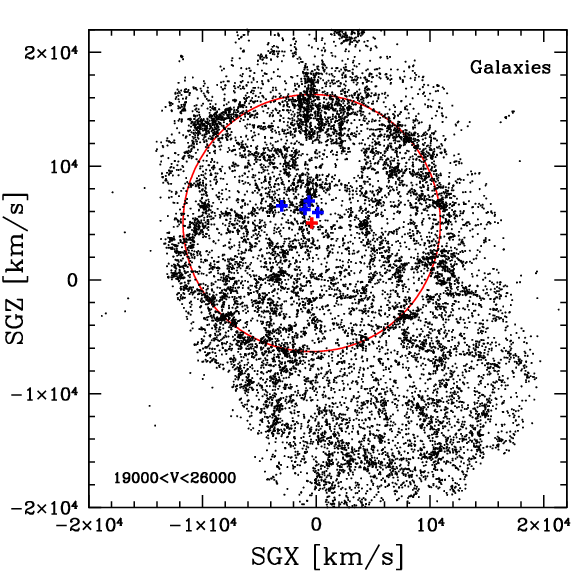}
\caption{
\textit{Top panel}: Supergalactic SGX$-$SGZ projection of all northern Cosmicflows-4 galaxies with $19000<V_{cmb}<26000$~\kms. \textit{Bottom panel}: Same as top panel with the addition in red of a circle of radius 11300 km/s centered at SGX$_c=-400$~\kms, SGZ$_c=5000$~\kms\ and blue crosses at the locations of Abell clusters A1781, A1795, A1825, and A1831. 
}
\label{fig:xzviews}
\end{figure}

Restricting the velocity range to the interval $19,000-26,000$~\kms, the domain including the Sloan Great Wall, a view from the third orthogonal direction emphasizes what appears to be a ring structure, shown in the top panel of Figure~\ref{fig:xzviews}.  A reasonable by-eye fit to the structure is given by the red ring of radius 11,300~\kms\ centered at SGX$_c= -400$~\kms, SGZ$_c= 5000$~\kms\ in the bottom panel, which we show later in this work is also statistically justified when considering the full 3-dimensional galaxy distribution.

This structure has been noted by \citet{Einasto+16} as the most prominent of several shell-like structures revealed in the main SDSS sample.  These authors looked for features around clusters and groups of varying richness and found the richer groups provided the more convincing evidence as the centers for shell-like structures. These authors considered but did not favor that any of the features they detect are related to the BAO.  As will be discussed further, we find contrary evidence that the ring seen in Figure~\ref{fig:xzviews} does indeed form part of a large coherent `BAO shell' --- the biggest contribution to the overall BAO signal that we will report. In any event, this apparent ring structure at a distance of $\sim 250$~Mpc from us is one of the largest structures observed in the nearby Universe to date and links together a number of hitherto disconnected components of our cosmic neighborhood.  We name this remarkable structure ``Ho'oleilana''.\footnote{``Sent murmurs of awakening'' from the Hawaiian Kumulipo creation chant: {\it Ho'oleilei ka lana a ka Po uliuli}, ``From deep darkness came murmurs of awakening''.}

\begin{figure}[!]
\centering
\includegraphics[width=.97\linewidth] {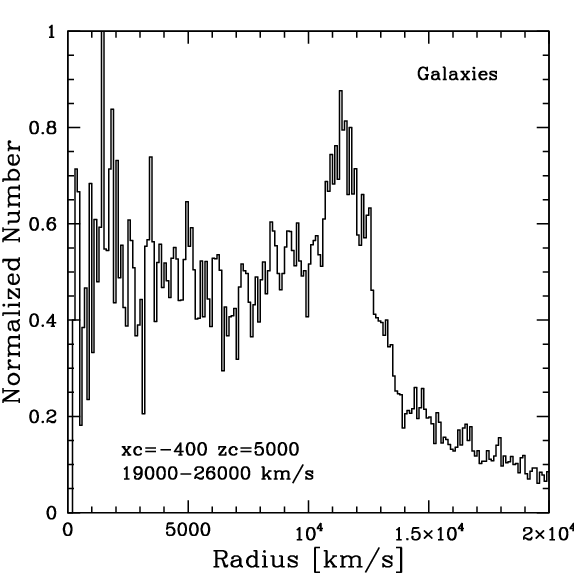}
\includegraphics[width=.97\linewidth]{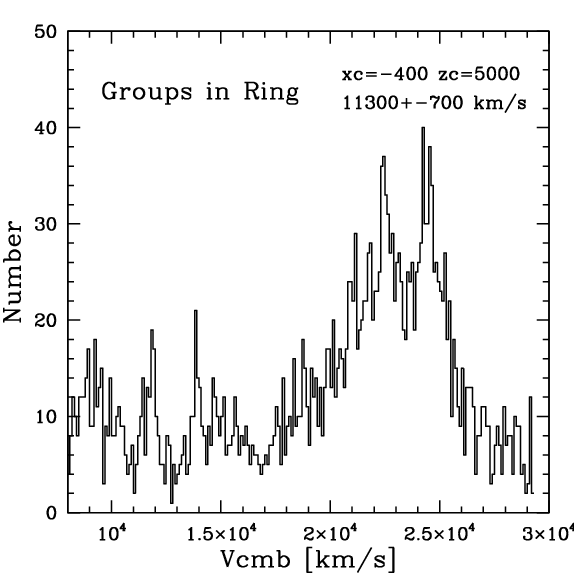}
\caption{
\textit{Top panel}: Histogram of numbers of galaxies in the systemic velocity window 19,000$-$26,000~\kms\ as a function of radius from a center at SGX$_c=-400$~\kms, SGZ$_c=5000$~\kms\ in 100~\kms\ bins. Numbers are normalized by the area within each 100~\kms\ bin.
\textit{Bottom panel}: Histogram of velocities in 100~\kms\ bins of groups that are projected into a ring of width with bounds 10,600$-$12,000~\kms\ centered at SGX$_c=-400$~\kms, SGZ$_c=5000$~\kms.
}
\label{fig:rv}
\end{figure}

The two histograms in Figure~\ref{fig:rv} give more insight into the properties of Ho'oleilana.   The upper panel gives the numbers of our galaxies in annular rings about the center illustrated in the bottom panel of Figure~\ref{fig:xzviews}, normalized by the area in each ring. There is an evident peak centered at 11,300~\kms. The lower panel shows the numbers of groups of galaxies projected into the ring with radius spanning 10,600$-$12,000~\kms\ as a function of systemic velocity in the CMB frame.  Numbers peak at $23,000\pm300$~\kms.\footnote{A comment regarding our plots sometime using all galaxies in our sample and sometimes using the galaxy groupings: statistics are better with all galaxies but ``finger of god'' velocity dispersions are suppressed with the compression into groups. There is consistency between group and all-galaxy presentations.} 

Several major features lie within this annulus.  The region of high density from 7~o'clock to 10~o'clock in Figure~\ref{fig:xzviews} is the main component of the Sloan Great Wall \citep{Gott+05}.  The Corona Borealis supercluster lies at $\sim12$~o'clock, the Ursa Majoris supercluster lies at $\sim4$~o'clock, and the Virgo-Coma supercluster lies at $\sim6:30$ \citep{Einasto+01}.  Lesser structures include SCL~95 at $\sim5$~o'clock and SCL~154 at $\sim10:30$. The Bo\"otes supercluster with 12 Abell clusters \citep{Einasto+01} is found to be close to, but not quite at, the center. This supercluster contains the Abell Richness 2 cluster A1795 at $V_{cmb}=19,204$~\kms\ plus three Abell Richness 1 clusters.  The projected positions of A1781, A1795, A1825, and A1831 are plotted in the lower panel of Figure~\ref{fig:xzviews}.

\section{Ho'oleilana in 3D}

\begin{figure}[!]
\centering
\includegraphics[width=.97\linewidth]{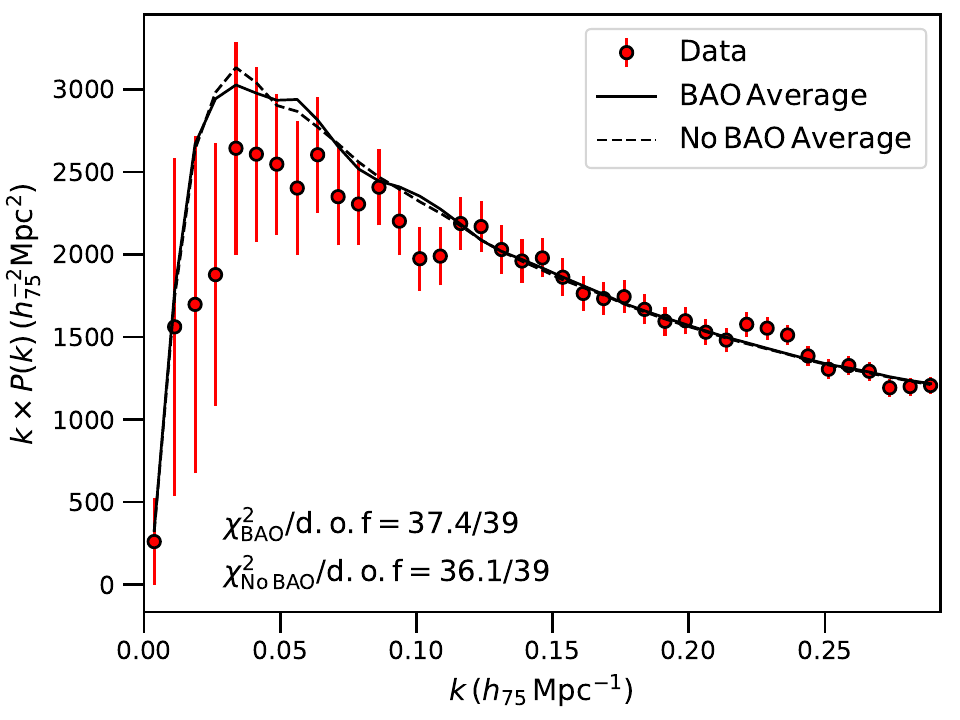}
\includegraphics[width=.97\linewidth]{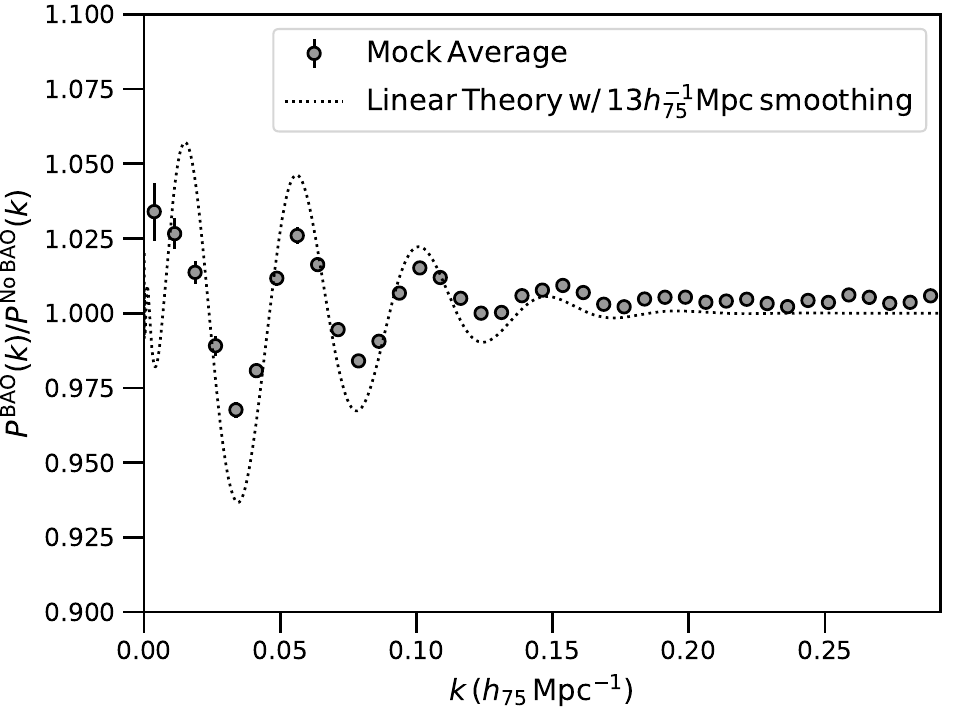}
\caption{
\textit{Top panel}: The power spectrum of the SDSS PV data compared to the average from sets of simulations with similar large-scale structure and selection effects, with and without the presence of BAO. Our method for populating the simulations with mock galaxies is tuned to this data, so we also report the $\chi^{2}$ difference for the two sets of simulations.
\textit{Bottom panel}: The average of the ratio of the power spectra for our BAO and no-BAO simulations. For reference, the black dotted line shows the linear theory model smoothed with a non-linear damping of 13 $h_{75}^{-1}$ Mpc (see Section~\ref{sec:baomodel}), which clearly demonstrates that the BAO is being suppressed in the no-BAO simulations. Note that the linear theory model has not been corrected for selection effects and galaxy bias so is not expected to match the amplitude of the data.
} 
\label{fig:pk}
\end{figure}

The BAO are expected to be spherical shells, rather than just the ring that has been identified above. The galaxies in the relevant region in the Cosmicflows-4 collection are overwhelmingly contributed by the SDSS Peculiar Velocity (SDSS PV) catalog \citep{Howlett+22}.  For three reasons the following statistical analysis of Ho'oleilana as a BAO feature is restricted to the elements of this catalog. This subset of the full Cosmicflows-4 collection has 1) a very well defined selection function; 2) a random, ungrouped catalog with the same selection function; and 3) an ensemble of $2\times256$ mock galaxy catalogs that also reproduce the large-scale structure, galaxy bias, and selection function of the data, but where in one half the BAO have been suppressed by generating the initial conditions of the simulations using a smoothed `no-wiggle' linear power spectrum \citep{Hinton+17}. Figure~\ref{fig:pk} shows the power spectra of the SDSS PV data and mocks, as well as the ratio of the two sets of simulations highlighting the BAO feature/suppression. These additional data products are essential for interpreting the significance of Ho'oleilana and fitting its size and shape, and only available for the SDSS PV portion of this catalog in the region of the Universe containing Ho'oleilana.

We explore the consistency of Ho'oleilana emerging from the physics of the BAO in two different ways: 1) by fitting the 3D radial distribution of galaxy groups in the SDSS Peculiar Velocity catalog with a physical model for the BAO feature; and 2) by blindly searching for similar contributors to the BAO signal in the simulations and data using the wavelet-convolution method of \cite{arnalte-mur+12}. We report on these tests in separate sections below. Overall, our findings provide evidence that Ho'oleilana is not a chance arrangement of galaxies, but instead a part of the total BAO signal in our nearby Universe.

\subsection{Modelling Ho'oleilana as a BAO feature}
\label{sec:HoBAO}
For our first test of the origin of Ho'oleilana, we begin by converting the SDSS PV catalogue from right ascension, declination and redshift to supergalactic cartesian coordinates assuming a fiducial flat Lambda Cold Dark Matter ($\Lambda$CDM) cosmological model with matter density $\Omega_{m}=0.31$ and expansion rate $H_{0}= 75\,h_{75}\mathrm{km\,s^{-1}\,Mpc^{-1}}$. We use the `group' redshifts from the data and simulations as this suppresses the impact of non-linear galaxy motions and is expected to make any large-scale coherent structures more prominent. 
Although each of the SDSS PV galaxies also has a measured distance from the Fundamental Plane, we do not use these for this analysis as their large uncertainties could `smear' out any apparent large-scale structures. Our use of `redshift-space' coordinates will do the same thing, but to a much lesser extent as the typical peculiar velocities of galaxies are much less than the typical distance uncertainties (the same reasoning is used when measuring the velocity clustering from such data; \citealt{Howlett19}). From there, we evaluate
\begin{equation}
N_{\mathrm{shell}}(r) = \frac{N_{D}(r)}{N_{D,tot}} \frac{N_{R,tot}}{N_{R}(r)} - 1
\end{equation}
in radial bins of width $\Delta r=5\,h_{75}^{-1}\mathrm{Mpc}$ centered on some location. $N_{D}(r)$ is the number of galaxies in each radial bin centered on $r$, while $N_{R}$ is the number of random, unclustered points. The subscript ``$\mathrm{tot}$'' denotes the total number of galaxies in the data and random catalogs, $N_{D,tot}=34,059$ and $N_{R,tot}=4\times10^{6}$ respectively. $N_{\mathrm{shell}}$ is hence normalised, such that for a completely homogeneous distribution of galaxies we expect $N_{\mathrm{shell}}(r) \equiv 0$. This normalization is verified by applying the same procedure to our simulations --- $\langle N_{\mathrm{shell}}(r) \rangle$, computed from 256 histograms of $N_{\mathrm{shell}}$ using the same central location and then averaged in each bin, is consistent with zero. This test indicates that the selection function (for instance, when approaching the edge of the available survey data) is correctly being mitigated by our method. The standard deviation in the 256 measurements in each bin is used as our uncertainty in the real data.

The strength and properties of any features in $N_{\mathrm{shell}}(r)$ will clearly depend on the assumed origin, so we explore a range of different possible central locations. Based on our initial finding of Ho'oleilana in the 2D distribution of data seen in Figure~\ref{fig:2views}, we choose a grid of 4225 nearby possible coordinates for the presumed center. In supergalactic X, Y and Z coordinates these span from $[-56, 208, 26]\,h_{75}^{-1}~\mathrm{Mpc}$ to $[8, 272, 90]\,h_{75}^{-1}~\mathrm{Mpc}$ in bins of width $[5, 2.5, 5]\,h_{75}^{-1}~\mathrm{Mpc}$ respectively.\footnote{The spacing of this grid is purposefully narrower in the SGY direction as in our preliminary work we found that the assumed central SGY coordinate was more correlated with the resulting constraints on the radius of Ho'oleilana, which arises from the fact that the most prominent aspect of Ho'oleilana is the 2D ring seen in the SGY plane of Figure~\ref{fig:xzviews}.}

\begin{figure}[!]
\centering
\includegraphics[width=.97\linewidth]{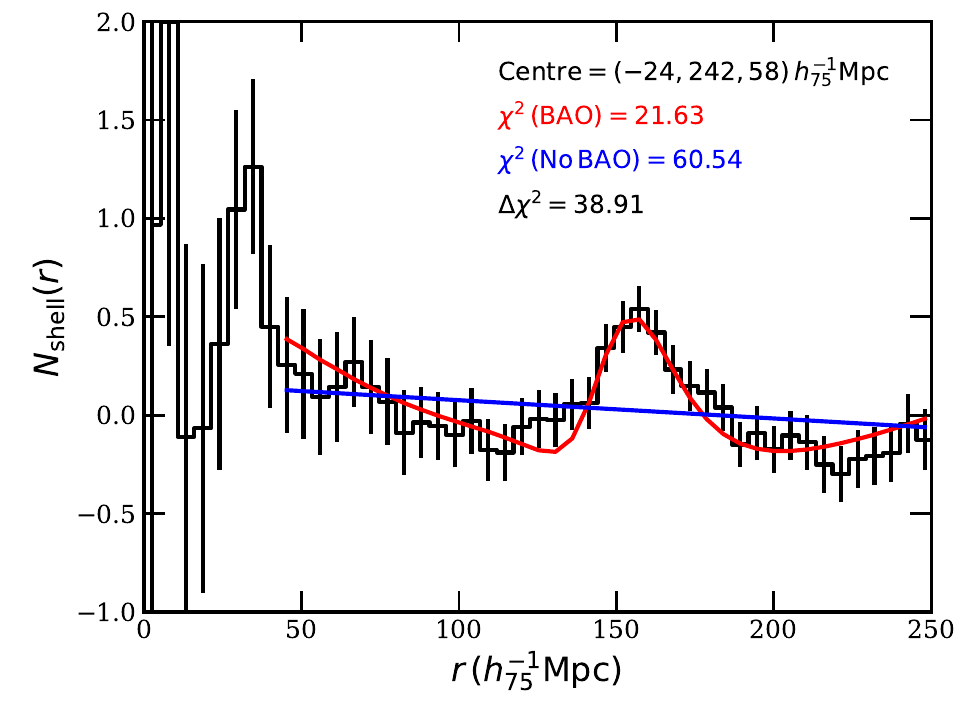}
\caption{A normalized histogram of the number of galaxies as a function of distance from the center of Ho'oleilana. Error bars are computed using simulations fully reproducing the selection function of the data. Ho'oleilana, at a distance $r\approx 150~ h^{-1}_{75}~\mathrm{Mpc}$ from the central location, is well fit by a physical BAO model (red line). Compared to the expectations for a random field of galaxies (blue line), Ho'oleilana is detected at greater than $6\sigma$ significance as shown by the relative difference in $\chi^{2}$ statistic.}
\label{fig:baomodel}
\end{figure}

Figure~\ref{fig:baomodel} shows the normalised radial distribution of groups at the central location where the supposed BAO `bump' is most prominent, $[\mathrm{SGX}, \mathrm{SGY}, \mathrm{SGZ}] = [-24, 242, 58]\,h^{-1}_{75}~\mathrm{Mpc}$. This same figure shows our best-fitting model of the feature which is used to define the detection significance as elaborated on below.

\subsubsection{BAO model}
\label{sec:baomodel}

The detection significance is derived from fitting a model with and without a BAO feature to $N_{\mathrm{shell}}$ and comparing the difference in best-fitting $\chi^{2}$ between the two. We generate our model based on the work of \cite{Eisenstein+98,Eisenstein+07,Slepian+16}, modelling the histogram of $N_{\mathrm{shell}}$ as a mass-profile in configuration space generated from fourier/Hankel transforming a smooth, `no-wiggle' transfer function describing the central overdensity ($T_{nw}(k)$) and a BAO `wiggle' only transfer function with non-linear damping ($T_{w}(k)$). We model this damping as an additional Gaussian smoothing with a free parameter $\Sigma_{nl}$ to account for evolution in the positions of galaxies since the time the BAO was frozen-in and $z\approx0.07$. 

Our model also needs to include flexibility in the radial distance of the feature from the central location, its amplitude (or equivalently, linear galaxy bias) and the possibility that the entire region we consider is over- or underdense compared to the full SDSS PV data. We do this by including three free parameters, $\alpha$, $B$ and $N_{0}$ for these three characteristics. Finally, in our tests we also found that the relative amplitudes of Ho'oleilana and the central overdensity could not be well described with a single amplitude parameter --- the measured values of $N_{\mathrm{shell}}$ for small $r$ and near the peak of Ho'oleilana itself are larger than would be predicted by just our linear theory model leading to a poor fit. We attribute this mainly to non-linear clustering and galaxy bias and to account for this effect we included an additional linear term $N_{1}$ on top of the constant $N_{0}$, slightly increasing the order of the polynomial used to marginalise over the overall shape of the mass profile. This same procedure is used in standard BAO fitting, where a cubic or quartic polynomial is typically used to marginalise over the shape of the correlation function rather than attempting to model non-linear effects, which ensures the constraints on the BAO peak position itself are insensitive to the broadband shape of the clustering (e.g., \citealt{Ross+15,Hinton+20,Alam+21}).

Overall, our BAO model hence contains five free parameters and can be summarised as
\begin{equation}
N^{\mathrm{model}}_{\mathrm{shell}}(r) = Br^{2}\int_{0}^{\infty} \frac{k^{2}dk}{2\pi^{2}} T(k)j_{0}(\alpha k r) + N_{1}r + N_{0},
\end{equation}
where
\begin{equation}
    T(k) = T_{w}(k)e^{-1/2k^{2}\Sigma_{nl}^{2}} + T_{nw}(k)
\end{equation}
and $j_{0}(x)$ is the zeroth-order spherical Bessel function. The model in the absence of BAO can be recognised from the above expressions as the case where we set $T(k) = T_{nw}(k)$ or equivalently, $\Sigma_{nl} \rightarrow \infty$ and fix $\alpha=1$. The model without BAO hence has three free parameters and should only reproduce the central overdensity.

Modelling the `wiggle' and `no-wiggle' transfer functions is not quite trivial. Firstly, one is required to assume a `template' cosmological model to create a BAO feature that can then be dilated by the value of $\alpha$. To allow for comparison/combination later with the \cite{Planck+20} constraints on the sound horizon, we adopt a template cosmology close to the \cite{Planck+20} best-fit. Parameter specifications are given in Table~\ref{tab:cosmo}. 
\begin{table}
    \centering
    \begin{tabular}{c|c|c}
       \hline
       \hline
       Parameter & Template & No Baryon Template  \\
       \hline
         $\Omega_{b}h^{2}$ & 0.0224 & 0.001\\
         $\Omega_{cdm}h^{2}$ & 0.1199 & 0.1412\\
         $H_{0}\,(\kmsMpc)$ & 67.51 & 67.51 \\
         $n_{s}$ & 0.9653 & 0.9653 \\
         $A_{s}$ & $2.25\times10^{-9}$ & $2.25\times10^{-9}$ \\
         $N_{eff}$ & 3.046 & 3.046 \\
         $\sum M_{\nu}\,(\mathrm{eV})$ & 0.06 & 0.06 \\
       \hline
    \end{tabular}
    \caption{A summary of the cosmological parameters used in creating our BAO model template, and for converting our constraints on the BAO model parameter $\alpha$ to cosmological constraints.}
    \label{tab:cosmo}
\end{table}

Note that this cosmology is not the same as that used to convert our catalog redshifts to distance (nor does it have to be), but it is important that the template cosmology be the one used to compute any cosmological constraints from our fit to $\alpha$. Secondly, the presence of baryons also introduces Silk damping effects in the transfer function, suppressing it on small scales, as well as adding the BAO. One cannot simply then take a numerical transfer function evaluated for a cosmology with and without baryons and difference the two to extract the wiggles. \cite{Eisenstein+98} provide fitting formulae for the smooth transfer function $T_{EH}(k)$ that could be used, although the approximations used therein for the sound horizon and equality scales are considered not quite accurate enough for modern BAO analyses \citep{Anderson+14}. Instead, we take a hybrid approach, and compute both numerical and \cite{Eisenstein+98} smooth transfer functions for our fiducial cosmology, and a second `no-baryon' cosmology (where $\Omega_{b}$ is reduced by a factor of $\sim20$ as also shown in Table~\ref{tab:cosmo}). We then compute the `no-wiggle' and `wiggle' transfer functions as
\begin{equation}
T_{nw}(k) = T_{EH}(k)\frac{T_{no\,baryon}(k)}{T_{EH,no\,baryon}(k)},  \quad
T_{w}(k) = T(k) - T_{nw}(k)
\end{equation}
which effectively corrects the \cite{Eisenstein+98} smooth transfer function $T_{EH}(k)$ so that its form is more representative of the broadband shape of the numerical transfer function before subtracting the two.

We perform our fits with both BAO and no BAO models restricting to scales $40\,h_{75}^{-1}~{\rm Mpc} < r < 250\,h_{75}^{-1}~{\rm Mpc}$, avoiding non-linearities at the core of the central overdensity. We perform a full MCMC fit using the \texttt{dynesty} sampler \citep{Speagle+20} \textit{at each} proposed center from Section~\ref{sec:HoBAO} and obtain the best-fit and errors on the five model parameters.

\subsubsection{Results}

From the $\chi^{2}$ difference between our two models, we conclude that Ho'oleilana, centered at the aforementioned coordinates, is detected at greater than $6\sigma$ significance compared to the expectations of a random field of galaxies, and there is excellent agreement between the data and physical BAO model. Of all the possible centers we tested, $716/4225$ and $127/4225$ result in greater than $4\sigma$ and $5\sigma$ detections respectively. We caution however that our choice of centers was conditioned on our visual identification of a feature in the data and so the relative fraction of $4$ and $5\sigma$ centers should not be interpreted in the usual way chance events are interpreted.

Of the five free parameters we fit for, $B$, $N_{0}$ and $N_{1}$ are nuisance parameters and so not commented on further in this work. For the others we perform an average of the posteriors at each proposed center with greater than $3.25\sigma$ significance (of which there are 1661). This threshold was chosen as it is close to the turnover point in a histogram of the BAO strengths across all the centers we tested. Combining fits in this way allows us to propagate the uncertainty in the central location into our constraints on the radius and damping within the BAO model. From just the fit using our most likely center for Ho'oleilana (see Fig.~\ref{fig:baomodel}), we find $\alpha = 0.87\pm0.01$, while averaging over all centers above our threshold gives $\alpha = 0.88^{+0.06}_{-0.09}$, indicating that a substantial portion of our error budget comes from the uncertainty in the central location.

Our combined chain also gives $\Sigma_{nl} = 12.8^{+0.8}_{-5.8}\,h_{75}^{-1}~\mathrm{Mpc}$. Under the condition that it is a remnant of the BAO, our constraint on $\Sigma_{nl}$ implies Ho'oleilana has undergone additional Gaussian smoothing due to the bulk motions of galaxies in the intervening years between $z\simeq1060$ (when the BAO feature was first frozen in) and the observed $z\approx0.07$, with standard deviation $\sim 13 \,h_{75}^{-1}\,\mathrm{Mpc}$. This value of $\Sigma_{nl}$ can be compared to the expected dispersion from linear theory. Given we are considering only relative separations from a fixed central point, rather than pairwise separations between objects, we expect this value to be comparable to the dispersion predicted to arise from the linear-scale velocities of galaxies. At $z=0$ for a $\Lambda$CDM cosmological model assuming General Relativity, this is given by
\begin{equation}
\frac{\sigma_{v}}{H_{0}} = \frac{\Omega_{m}^{0.55}}{\sqrt{6}\pi}\biggl(\int_{0}^{\infty} P_{\mathrm{lin}}(k) dk\biggl)^{1/2} \approx 4.2\,h_{75}^{-1}~\mathrm{Mpc}, 
\end{equation}
where the latter approximation is obtained using our template cosmology.\footnote{Evaluating this expression at the proper redshift of $z\approx0.07$ gives the same answer to within the precision quoted here.} One would expect non-linear evolution and redshift-space distortions to increase this value somewhat. We consider our measurement consistent with our expectations of the typical distances traversed by individual galaxies from their formation to the present day. 
Our recovered value for $\alpha$ can also be further interpreted, but this is done in Section~\ref{sec:cosmo}.

\subsection{Prevalence in simulations}

To quantify the true significance of Ho'oleilana without conditioning on our visual identification we make use of our simulations with and without BAO. The key questions are 1) whether Ho'oleilana could have been identified without an \textit{a priori} knowledge of such a structure in our nearby Universe; and 2) to what extent would we find similar features in simulations with large-scale structure, but where we know there are no BAO.

To answer these questions, we turn to a variant of the algorithm developed in \cite{arnalte-mur+12}, implemented in the publicly available BAO `centerfinder' code of \cite{Brown+21}. Our algorithm first uses a set of data and the random catalog to compute the overdensity field on a grid of cell size $5\,h_{75}^{-1}\,\mathrm{Mpc}$. It then convolves this field with a `BAO-wavelet'; a normalised, spherically-symmetric function that has a shape expected of a BAO feature. The wavelet has two free parameters, $r_{\mathrm{BAO}}$ and $s_{\mathrm{BAO}}$, which set the radius and width of the BAO feature. The result is, for each cell in the gridded overdensity field, a weight $W_{r,s}$ that describes its likelihood as the center of a BAO-like feature with the given radius and width. 

In order to determine whether or not a given field evidences the presence of BAO, \cite{arnalte-mur+12} proposed to average the values of $W_{r,s}$ over a subset of likely central locations (in their case halos, galaxy groups or galaxies likely to be found in the centers of large clusters), to find $B_{r,s} = \sum_{N_{\mathrm{subset}}} W_{r,s} /N_{\mathrm{subset}}$, claiming a positive value for this coefficient indicates a positive detection of BAO. We do the same, computing $W_{r,s}$ for each cell on our grid that contains at least one galaxy. We then construct $B_{r,s}$ by summing over all these cells (weighted by the number of galaxies in that cell). 

\begin{figure}[!]
\centering
\includegraphics[width=.93\linewidth,trim=0mm 0mm 0mm 10mm,clip]{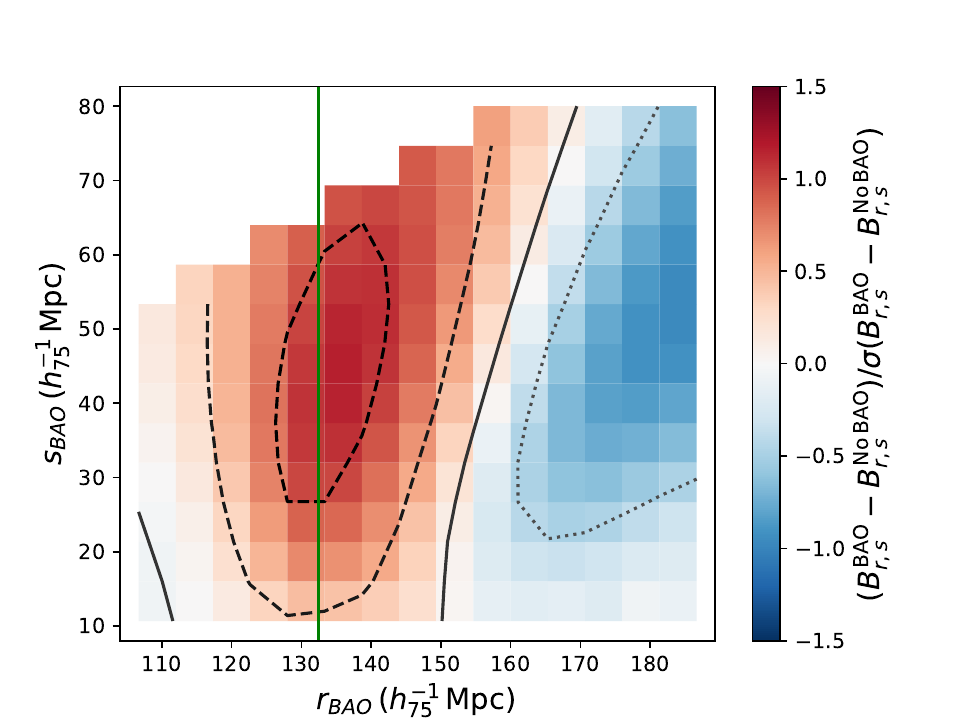}\\
\includegraphics[width=.93\linewidth,trim=0mm 0mm 0mm 10mm,clip]{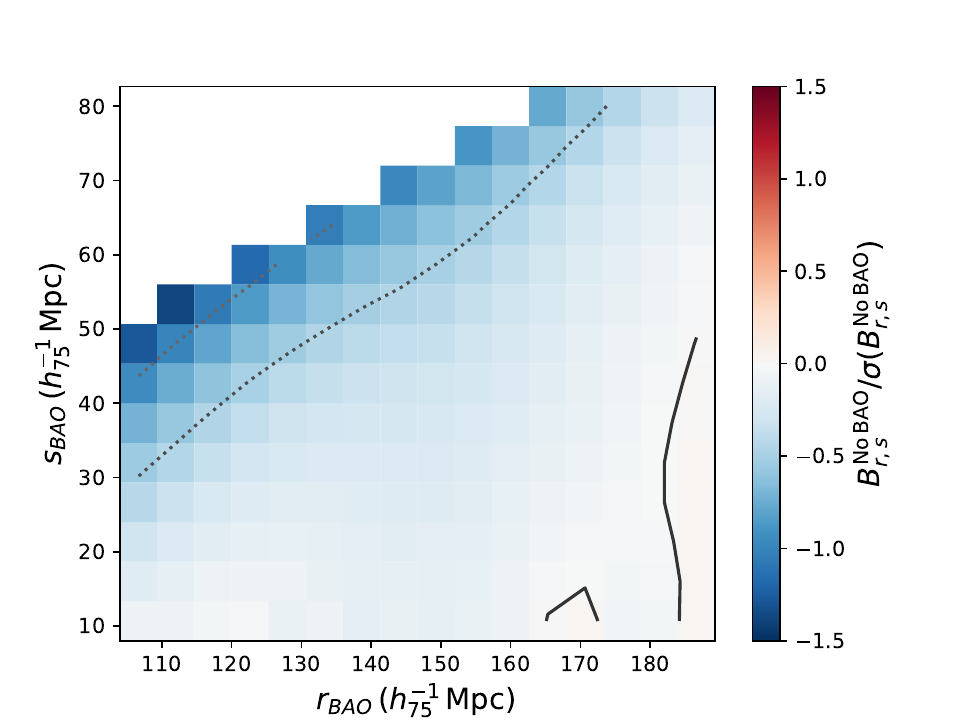}\\
\includegraphics[width=.93\linewidth,trim=0mm 0mm 0mm 10mm,clip]{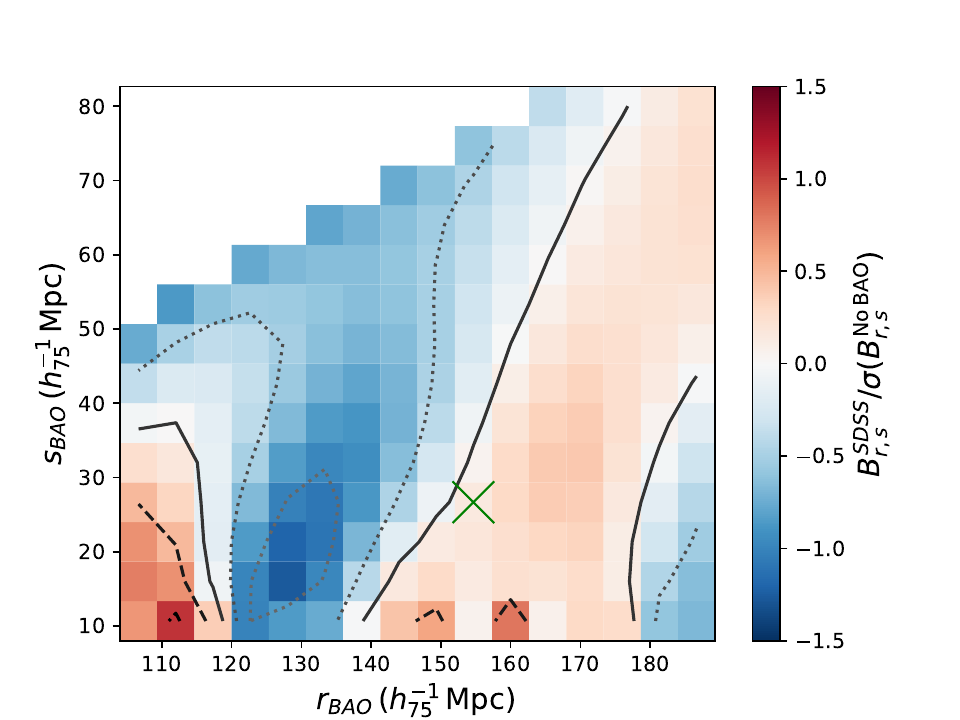}

\caption{BAO detection using the BAO-wavelet with different radii ($r_{\mathrm{BAO}}$) and widths ($s_{\mathrm{BAO}}$). The top panel of this figure shows the average values of the differences in the detection coefficient $B_{r,s}$ between the BAO and no-BAO simulations weighted by the variance. The vertical green line is the expected BAO radius in the mocks. The middle shows the average over the variance for only the no-BAO simulations. The bottom panel shows the same applied to the data. In all cases, the contours simply follow the color map, where the solid contour corresponds to a value of zero while the dashed (dotted) lines correspond to positive (negative) $0.5$ and $1.0$ sigma values. Dashed lines/red regions hence indicate values of the wavelet parameters with stronger BAO-like features. Finally, the green cross in the bottom panel indicates the wavelet parameters with the single largest BAO weight $W_{r,s}$, which coincides closely with the center, width and radius of Ho'oleilana derived via other means.}
\label{fig:brs}
\end{figure}

Figure~\ref{fig:brs} shows the results of this procedure for a wide variety of wavelet radii and widths. There is evidence that, on average, positive values of $B_{r,s}$ can be found preferentially in the BAO simulations at a scale around $r_{\mathrm{BAO}} = 135 \,h_{75}^{-1}\,\mathrm{Mpc}$. This corresponds well with the expected BAO radius given the cosmology used to generate the simulations ($132.4 \,h_{75}^{-1}\,\mathrm{Mpc}$), and indicates that the algorithm \textit{can} be used to identify BAO and the main contributors to the BAO. However, it is worth highlighting that the significance is relatively weak --- in an ideal scenario one would use a highly complete sample of galaxies to compute $W_{r,s}$ before summing this quantity only over objects close to the expected BAO centers. However, we suspect that our use of the SDSS PV sample (consisting only of ellipticals likely to be found in dense clusters) weakens the fluctuations in $W_{r,s}$ for larger radii and hence also weakens the signal in $B_{r,s}$.

The last panel in Figure~\ref{fig:brs} shows the same algorithm applied to the data. There is an excess of $B_{r,s}$ at somewhat larger radii around $r_{\mathrm{BAO}} = 160\,h_{75}^{-1}\,\mathrm{Mpc}$, although this result is also only weakly significant --- it is possible to find similar values of $B_{r,s}$ at these radii and widths even within simulations without BAO. In this same plot we identify with a green cross the values $r_{\mathrm{BAO}} = 155\,h_{75}^{-1}\,\mathrm{Mpc}$ and $s_{\mathrm{BAO}} = 27\,h_{75}^{-1}\,\mathrm{Mpc}$, which result in the largest single value for $W_{r,s}$ for all BAO-wavelets and grid cells we consider. This wavelet also returns a positive value of $B_{r,s}$ --- which is not guaranteed, given $B_{r,s}$ is the mean across all $\sim23,000$ cells in our overdensity grid.

\begin{figure}[!]
\centering
\includegraphics[width=.97\linewidth]{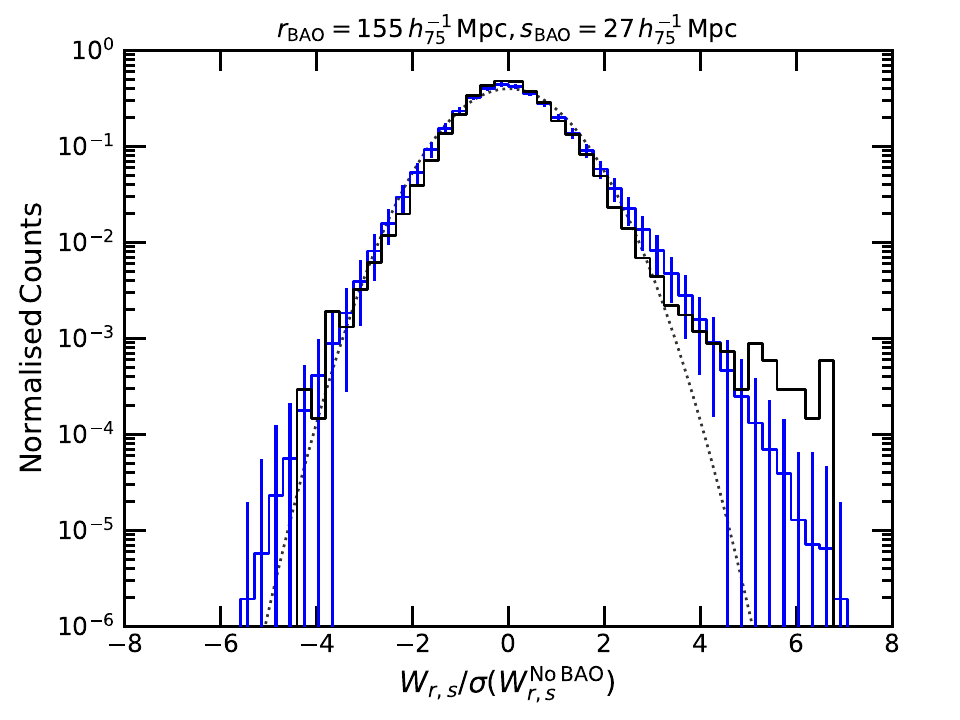}\\

\caption{A histogram of the weights $W_{r,s}$ for a BAO-wavelet with radius $r_{\mathrm{BAO}} = 155\,h_{75}^{-1}\,\mathrm{Mpc}$ and width $s_{\mathrm{BAO}} = 27\,h_{75}^{-1}\,\mathrm{Mpc}$. The weights are normalised by the standard deviation across the full sample, and positive values correspond to grid locations that demonstrate an overabundance of galaxies distributed within a BAO-like shell. There is a clear excess of positive weights in the data (black) compared to the average and standard deviation of the no-BAO mocks (blue). The dotted line shows a Gaussian distribution --- both the BAO and no-BAO mocks display a non-Gaussian tail of large overdensities. The largest positive weight corresponds to Ho'oleilana.}
\label{fig:wrs}
\end{figure}

Figure~\ref{fig:wrs} shows a histogram of all the values of $W_{r,s}$ we return (i.e., the value for each grid cell in our catalog containing at least one galaxy), for this choice of wavelet, for both the data and no-BAO simulations. There is a significant excess of grid cells with large weights. The most prominent excess corresponds closely to the center of Ho'oleilana identified in Section~\ref{sec:baomodel}. Our conclusion is that, using the independent algorithm of \cite{arnalte-mur+12} and without conditioning on our \textit{a priori} visual inspection, there is evidence of BAO-like features in the SDSS PV catalog, and that the strongest contributor to these features is Ho'oleilana.

To answer the second of our proposed questions, we similarly identify the largest single BAO-like feature (the grid cell with the largest $W_{r,s}$) within our 256 no-BAO simulations. We then generate radial profiles $N_{\mathrm{shell}}(r)$ centered at each of these 256 grid centres and fit the apparent BAO feature using our model from Section~\ref{sec:baomodel}. The aim is to compare the strength of these BAO-like features (in simulations which we know have heavily suppressed primordial BAO) to that inferred for Ho'oleilana.

\begin{figure}[!]
\centering
\includegraphics[width=.97\linewidth]{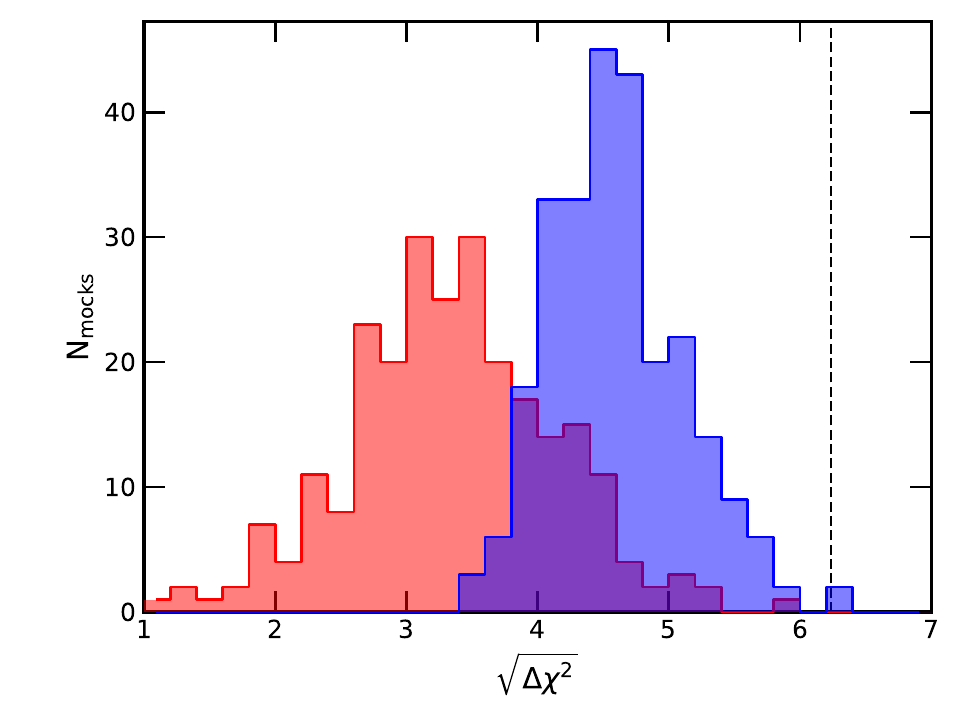}\\

\caption{Histograms of the $\chi^{2}$ difference between models for the radial distributions of galaxies with and without the BAO feature fit to the central location with the largest value of $W_{r,s}$ in each no-BAO mock. The dashed line shows the value from fitting Ho’oleilana in Section~\ref{sec:baomodel}. The blue histogram shows results allowing for any value of $r_{\mathrm{BAO}}$ and $s_{\mathrm{BAO}}$. The red histogram is conditioned on $r_{\mathrm{BAO}}=155\,h_{75}^{-1}\,\mathrm{Mpc}$ and $s_{\mathrm{BAO}}=27\,h_{75}^{-1}\,\mathrm{Mpc}$, the radius and width which returns the largest value of $W_{r,s}$ in the data, corresponding Ho'oleilana.}
\label{fig:chi2hist}
\end{figure}

Figure~\ref{fig:chi2hist} shows this comparison, as histograms of the $\chi^{2}$ difference between the BAO model and a straight line fit for each of the no-BAO mocks, in the first instance, allowing for freedom in the values of $r_{\mathrm{BAO}}$ and  $s_{\mathrm{BAO}}$ and, then, restricting to the radius and scale $r_{\mathrm{BAO}}=155\,h_{75}^{-1}\,\mathrm{Mpc}$ and $s_{\mathrm{BAO}}=27\,h_{75}^{-1}\,\mathrm{Mpc}$ identified previously. Of the 256 simulations we test, two of them have features of any radius or width that are more significant than Ho'oleilana. We can hence conclude that the probability of Ho'oleilana being a chance alignment, and not associated with the primordial BAO, is $<1\%$. None of the simulations we test have a feature with the same radius and width as Ho'oleilana that is as significant. These tests provide reasonably strong evidence that Ho'oleilana is \textit{not} a chance occurrence, although the possibility cannot be ruled out completely.

\section{Cosmological constraints}
\label{sec:cosmo}

Under the assumption that Ho'oleilana is a significant contributor to the true, full, BAO signal, and that its properties are representative of the translationally-averaged BAO, we can use its radius to extract cosmological information. The parameter $\alpha$ fit in Section~\ref{sec:baomodel} is proportional to the ratio between the size and distance to the BAO, and so encodes this information. From \cite{Eisenstein+05}
\begin{equation}
\alpha = \frac{D_{v}}{r_{\mathrm{drag}}}\frac{r^{\mathrm{fid}}_{\mathrm{drag}}}{D^{\mathrm{fid}}_{v}},
\end{equation}
where $D_{v}$ is a measure of the distance to the center of the BAO, and $r_{\mathrm{drag}}$ is its radius, or more formally, the size of the sound horizon at the baryon-drag epoch after the photons and baryons have decoupled and the baryon inertia has subsided. The superscript ``fid'' denotes these quantities in our fiducial cosmology. $D_{v}$ is the `volume averaged' distance, which is a hybrid of two parts angular diameter distance $D_{A}$, and one part Hubble parameter $H(z)$ arising from the spherical symmetry of the BAO,
\begin{equation}
\frac{D_{v}(z)}{r_{\mathrm{drag}}} = \biggl(\frac{cz}{H(z)r_{\mathrm{drag}}}\biggl)^{1/3}\biggl(\frac{1+z}{2\sin^{-1}(r_{\mathrm{drag}}/2D_{A}(z))}\biggl)^{2/3}.
\label{eq:dv}
\end{equation}
This expression differs slightly to the conventional one \citep{Eisenstein+05} as we have avoided using the small-angle approximation given the low redshift of Ho'oleilana. 

\begin{figure}[!]
\centering
\includegraphics[width=0.83\linewidth]{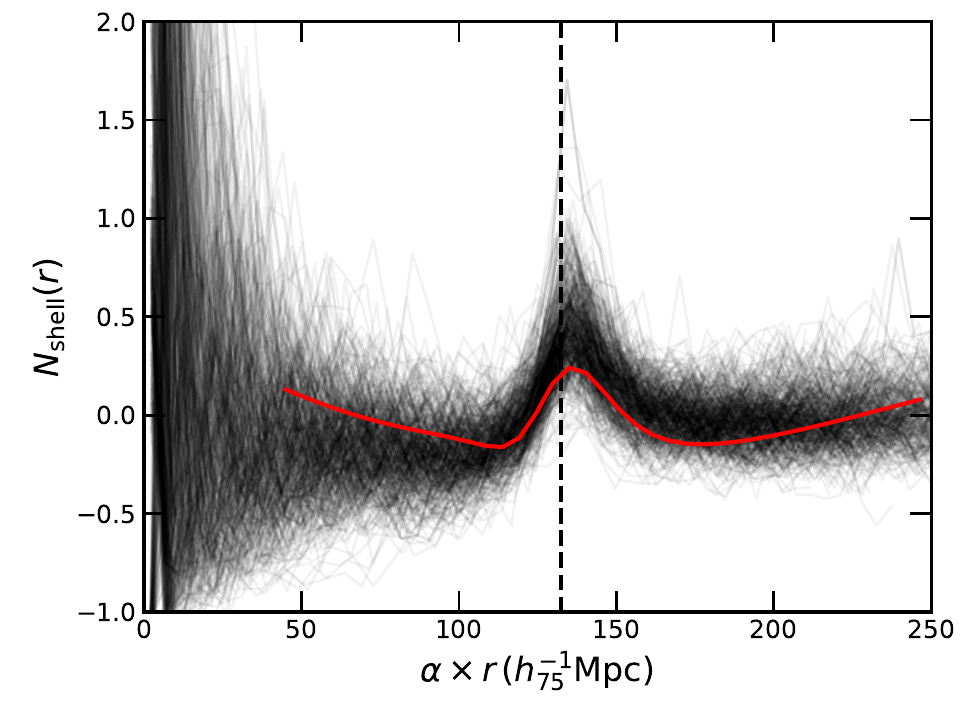}
\includegraphics[width=0.83\linewidth]{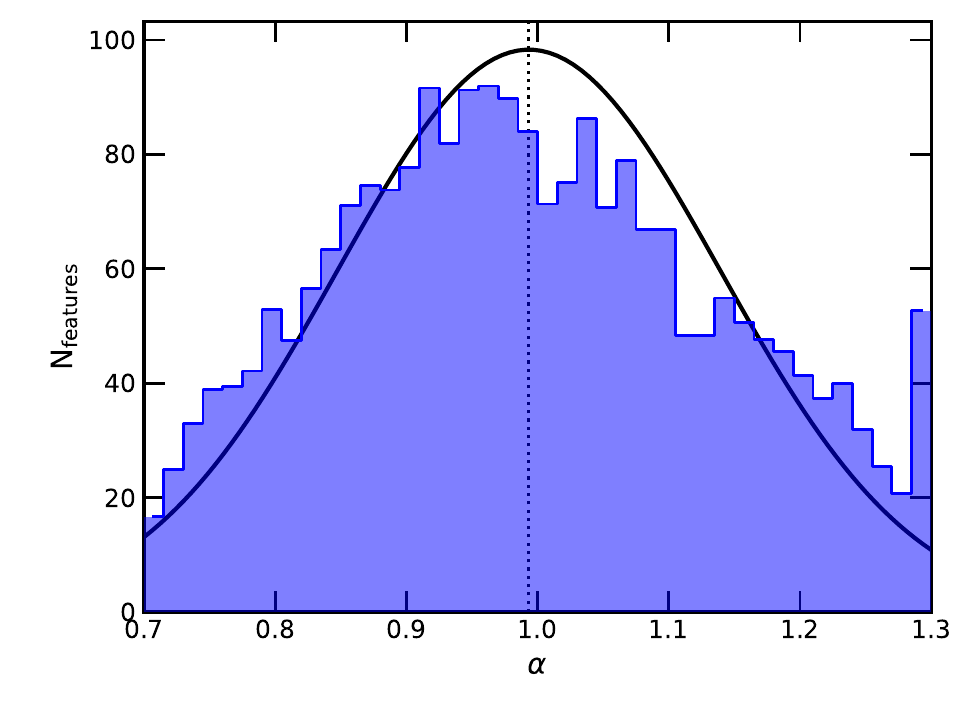}\\
\includegraphics[width=0.83\linewidth]{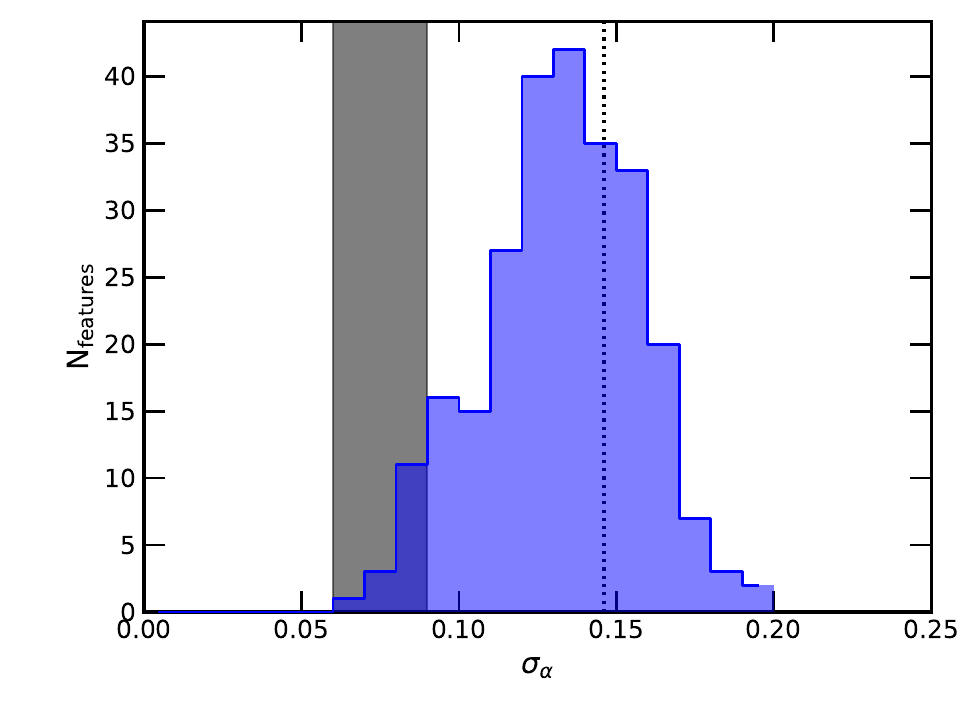}\\
\caption{\textit{Top panel}: A composite of single BAO features found in the 256 simulations we explore, restricting to simulations with positive BAO detections $B_{r,s}>0$, where we have scaled the radius of the BAO shell by the parameter $\alpha$ to align them for clarity. The red line shows the BAO model averaged over each of the fits to the individual detections. The dashed vertical line is the expected BAO radius given the input cosmology of the simulation. \textit{Middle panel}: A histogram of the recovered $\alpha$ values from fitting these single BAO features with the model from Section~\ref{sec:baomodel}, where the solid line is a Gaussian fit to the distribution with mean given by the dotted line.
\textit{Bottom panel}: A histogram of the standard deviation in best-fit $\alpha$ measured from features \textit{within} each single mock (i.e., we measure 256 values of the standard deviation from our 256 mocks.). The black shaded region shows our baseline error on $\alpha$ from Ho'oleilana, while the dashed line is the standard deviation measured from across all features and mocks (i.e., all the measurements shown in the middle panel).}
\label{fig:baomodelcomp}
\end{figure}

To test the validity of our approach, we first perform a similar analysis on our simulated catalogs with BAO. As was done for the no-BAO simulations, we identify the highest weighted central locations across the same range of $r_{\mathrm{BAO}}$ and $s_{\mathrm{BAO}}$ used previously in each of our 256 simulations (i.e., in each cell of Figure~\ref{fig:brs}). We then measure $N_{\mathrm{shell}}$ for each center and fit BAO and no-BAO models. Figure~\ref{fig:baomodelcomp} shows a composite of a random subset of our detected features in simulations with $B_{r,s}>0$, scaled by the best-fit value of $\alpha$ to make the BAO more prominent. We also plot the average best-fit BAO model and expected BAO radius given the input cosmology of the simulation. 

In addition, in Figure~\ref{fig:baomodelcomp} we also show histogram of these BAO fits, again restricting to mocks with $B_{r,s}>0$, but also normalising by the number of $s_{\mathrm{BAO}}$ bins at each $r_{\mathrm{BAO}}$.\footnote{This is important, because the requirements of the BAO wavelet to have $r_{\mathrm{BAO}}\ge 2s_{\mathrm{BAO}}$ means our list of the strongest BAO contributions at each combination of $r_{\mathrm{BAO}}$ and $s_{\mathrm{BAO}}$ contains more features with larger $r_{\mathrm{BAO}}$ and hence smaller $\alpha$. Normalising the histogram of $\alpha$ values by the number of $s_{\mathrm{BAO}}$ bins corrects for this selection bias.} The average $\alpha$ value is close to 1, indicating no bias in our method for identifying and fitting the individual BAO contributors. However, the uncertainty from the mocks both using only detections within a single mock realisation, or compared across realisations is slightly larger than that found from our fit to Ho'oleilana. This may simply be a result of the fact that Ho'oleilana is an exceptionally strong feature, even within mocks containing BAO. Or it may indicate an additional contribution that should be included in our error budget from sample variance. We hence elect to provide constraints and draw conclusions using both our fitted error on $\alpha$ from Ho'oleilana ($\alpha = 0.88^{+0.06}_{-0.09}$), and using the uncertainty derived from the scatter in the mocks ($\alpha = 0.88\pm0.14$) but fixed to our most likely center for Ho'oleilana. 

\begin{figure}[!]
\centering
\includegraphics[width=1.05\linewidth]{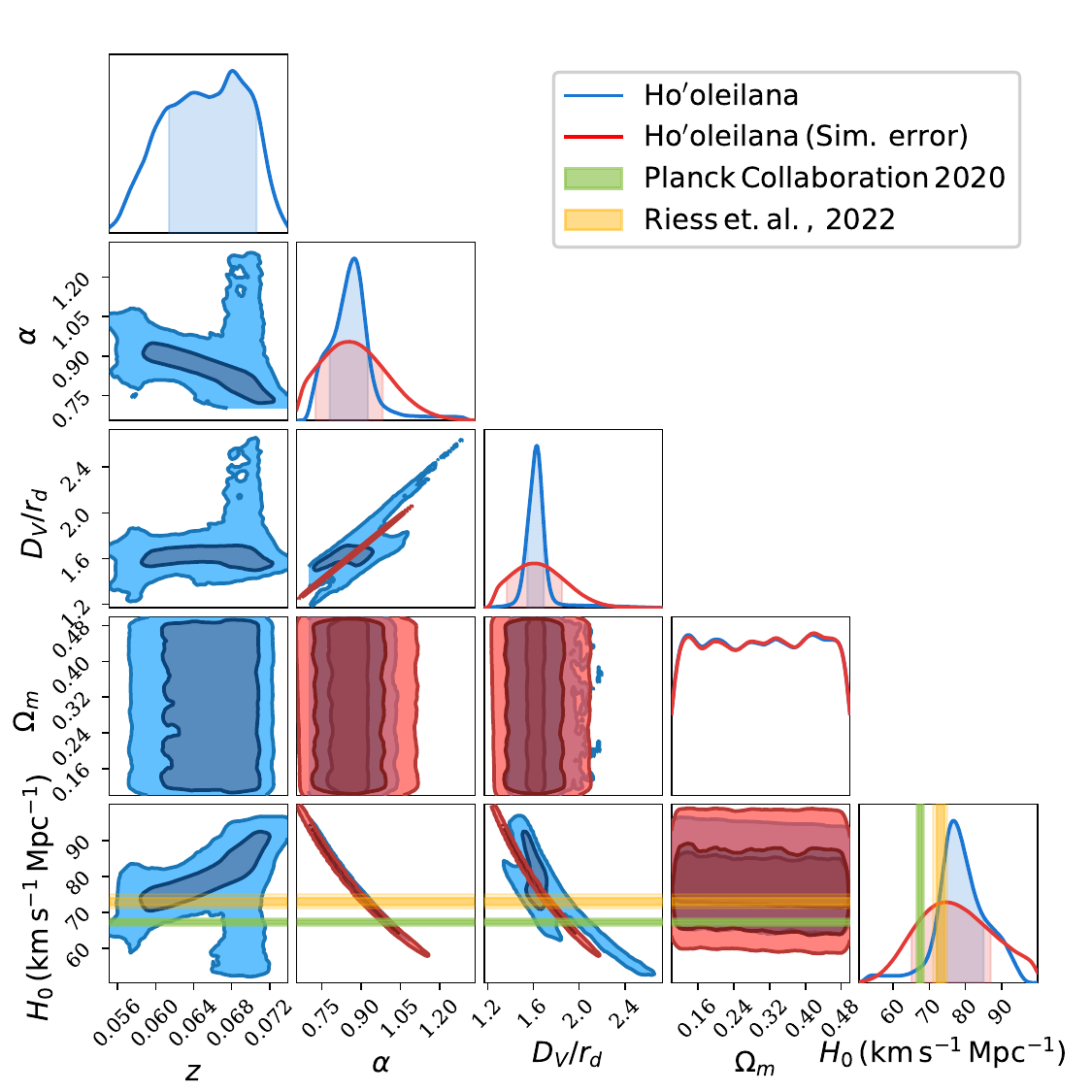}
\caption{Cosmological constraints from fitting Ho'oleilana with a BAO model using the uncertainty from Ho'oleilana alone allowing for uncertainty in the central location (blue) or using the the scatter seen in BAO simulations as the uncertainty fixing to our most likely center for Ho'oleilana (red). $\alpha$ is our BAO scaling parameter, $D_{v}/r_{\mathrm{drag}}$ is the ratio between the distance to the center of Ho'oleilana and its size, while $z$ is the redshift to its center. From these pieces of information, and assuming constraints on $r_{\mathrm{drag}}$ from early Universe physics, we obtain constraints on the matter content $\Omega_{m}$ and present day expansion rate of the Universe $H_{0}$ that favor other local direct measurements (orange band) rather than that propagated from models of the early universe (green band). 
}
\label{fig:h0}
\end{figure}

Turning back to the data, we convert our values of $\alpha$ to a distance ratio $D_{v}/r_{\mathrm{drag}}$ using our fiducial cosmology. A substantial fraction of our uncertainty on $\alpha$ comes from marginalising over the possible centers --- similarly we then have a range of possible fiducial distances to the center of Ho'oleilana, which translates to uncertain values for our redshift and fiducial distance ratio of $z=0.068_{-0.007}^{+0.003}$ and $D_{v}^{\mathrm{fid}}/r_{\mathrm{drag}}^{\mathrm{fid}} = 1.99_{-0.20}^{+0.07}$ respectively. Propagating this constraint properly alongside the constraint on $\alpha$ we recover $D_{v}/r_{\mathrm{drag}} = 1.63_{-0.08}^{+0.07}$. \footnote{Note that this propagation is done by converting each point of our MCMC chains --- one cannot simply multiply the reported constraints on $\alpha$ and $D_{v}^{\mathrm{fid}}/r_{\mathrm{drag}}^{\mathrm{fid}}$ because these are extremely correlated, see Figure~\ref{fig:h0}.}. In the case where we fix the central location of Ho'oleilana but use the mock scatter as our error on $\alpha$, we have a fixed $D_{v}^{\mathrm{fid}}/r_{\mathrm{drag}}^{\mathrm{fid}}=1.88$ and find $D_{v}/r_{\mathrm{drag}} = 1.66\pm0.26$.

\begin{figure}[!]
\centering
\includegraphics[width=0.97\linewidth]{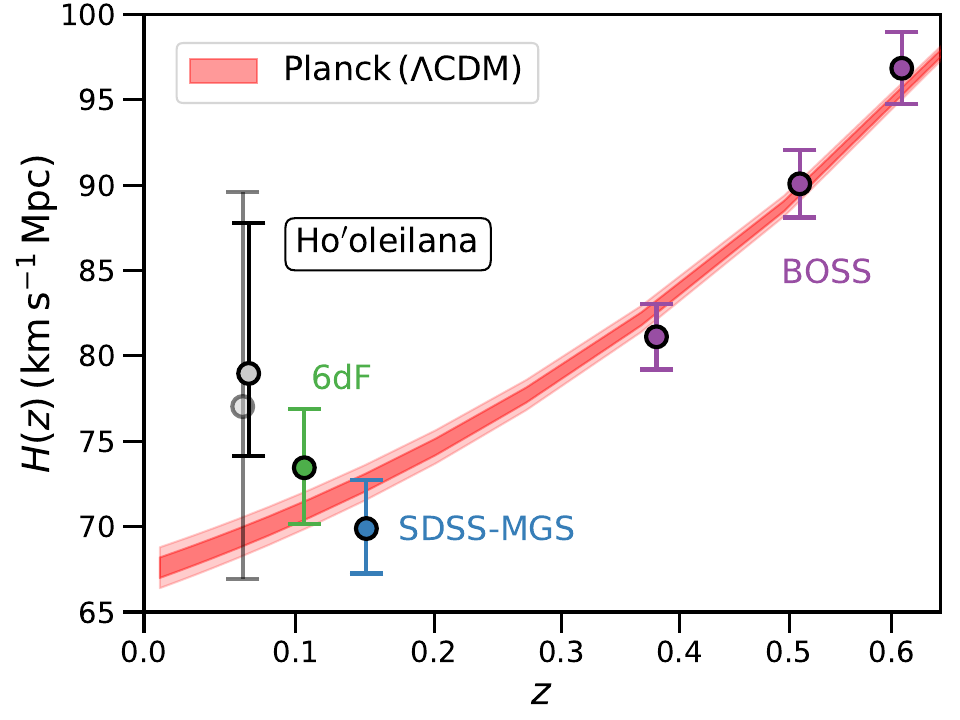}
\caption{A comparison of the expansion rate of the Universe as a function of redshift from Ho'oleilana compared to other \textit{statistical} BAO measurements \citep{Beutler+11,Ross+15,Alam+17}, assuming a prior on the BAO size from early Universe physics. Bright and faint points represent our measurements using the uncertainty from Ho'oleilana alone, or taking the scatter from BAO simulations as the error, respectively. The solid band shows the predicted expansion rate for a $\Lambda$CDM cosmological model from \citet{Planck+20}.}
\label{fig:H_z}
\end{figure}

Finally, we adopt a prior on the sound horizon $r_{\mathrm{drag}} = 147.13\pm0.26\,\mathrm{Mpc}$ from \citet{Planck+20}, and fit our combined posterior for $\Omega_{m}$ and $H_{0}$. Although we have effectively allowed the redshift to vary in incorporating all our proposed centers, the general low redshift of Ho'oleilana makes it an almost pure probe of $H_{0}$, with little constraining power on $\Omega_{m}$. This is reflected in our final constraints, shown in Figure~\ref{fig:h0}, where $\Omega_{m}$ is unconstrained by our methodology. We find $H_{0} = 76.9^{+8.2}_{-4.8}$ \kmsMpc\ and $H_{0} = 74.7^{+12.4}_{-9.7}$ \kmsMpc\ using the uncertainties of $\alpha$ from Ho'oleilana only and from the distribution of our simulations respectively. For the former, the majority of the uncertainty on $H_{0}$ comes from the expected variance in $N_{\mathrm{shell}}$ as measured from the mocks (reflected in the error bars in Fig.~\ref{fig:baomodel}) and our uncertainty in the central location/redshift. Our constraints on the expansion rate of the Universe, and comparisons to other measurements, including statistical BAO, are shown in Figures~\ref{fig:h0} and~\ref{fig:H_z}. Being a single feature rather than a statistical average, the errors bars from Ho'oleilana are larger than those from other large-scale structure surveys, however are still constraining enough to provide a preference for larger expansion rates.

Given the presence of Ho'oleilana, an interesting question is whether the clustering of the full SDSS PV sample contains any evidence of \textit{statistical} BAO. The power spectrum measurements presented in Fig.~\ref{fig:pk} were analysed using the BAO fitting code Barry \citep{Hinton+20} taking into account the survey window function, however, unfortunately, we found no significant statistical BAO detection. Nonetheless, statistical BAO have been detected in both the 6dF and SDSS Main galaxy surveys \citep{Beutler+11,Ross+15}, which cover an effective volume only a few times larger than the catalogue we analyse here. The latter of these also covers the redshift range $0.07 < z < 0.20$ and so partially overlaps with our sample. It is worth noting that the SDSS results of \cite{Ross+15} relied on BAO reconstruction to enable their detection. As such, further investigation of the SDSS PV sample, using both the correlation function and BAO reconstruction, is warranted.

\section{Discussion and Summary}

Although Ho'oleilana was identified as a two-dimensional feature, a significant component of the signal comes from the foreground part of the three-dimensional shell. (The back side far edge falls slightly beyond the $z=0.1$ limit of our sample.) Remarkably, the Coma Cluster, the Center for Astrophysics Great Wall \citep{deLapparent+86}, and structure coursing up to the Hercules supercluster \citep{Einasto+01, shapley34} lie along the foreground surface of the posited BAO phenomenon.  The famous Bo\"otes Void \citep{kirshner81} lies within the embrace of Ho'oleilana. 
Near the center of Ho'oleilana is the Bo\"otes supercluster \citep{Einasto+01}, presumed to be the manifestation of the matter concentration that gave birth to the BAO \citep{Weinberg+13}.  In detail, the domain of the central supercluster about the dominant A1795 is diffused over $\sim 50 h_{75}^{-1}$~Mpc around the geometric center of the BAO shell

\begin{figure}[!]
\centering
\setlength{\fboxsep}{0pt}\fbox{\includegraphics[width=1.0\linewidth]{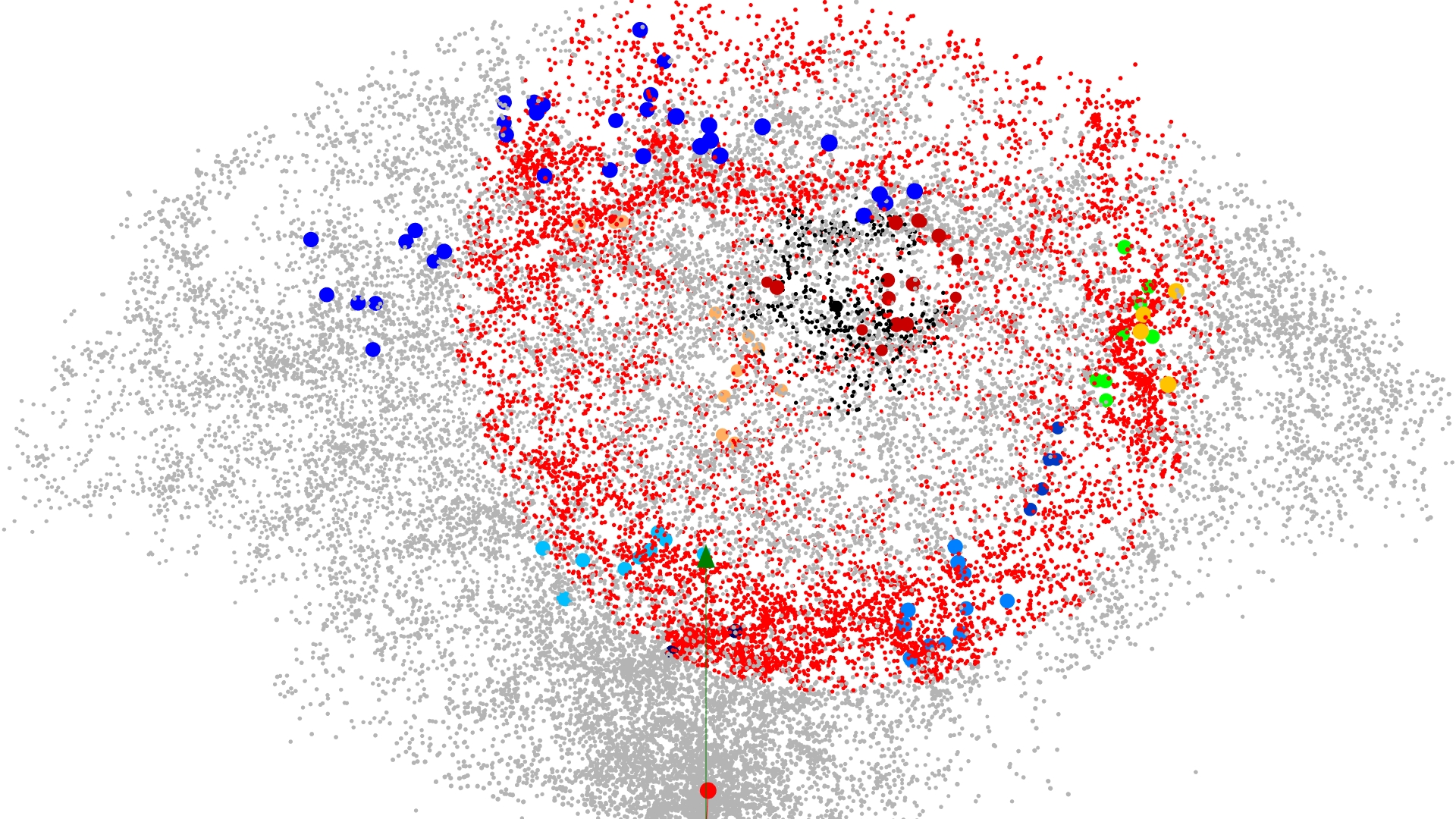}}
\caption{\href{https://vimeo.com/814958164/37504df76e}{Video visualization of the cosmography of Ho'oleilana}.  All objects in the north galactic hemisphere of the Cosmicflows-4 collection of galaxy groups are seen as points in gray while those lying within the shell of Ho'oleilana, of radius 11,492~\kms\ and width $\pm 2w$ where $w=837$~\kms\, are highlighted in red.  Major components in proximity to the shell are highlighted and identified by name.  The Bo\"otes supercluster lies near the center of Ho'oleilana and the Bo\"otes void lies interior to the shell structure.  Our home location is at the origin of the red, green, blue axes.  These axes have lengths 10,000~\kms\ and are directed toward positive SGX, SGY, SGZ respectively.}
\label{fig:video}
\end{figure}

\begin{figure}[!]
\centering
\setlength{\fboxsep}{0pt}\fbox{\includegraphics[width=1.0\linewidth]{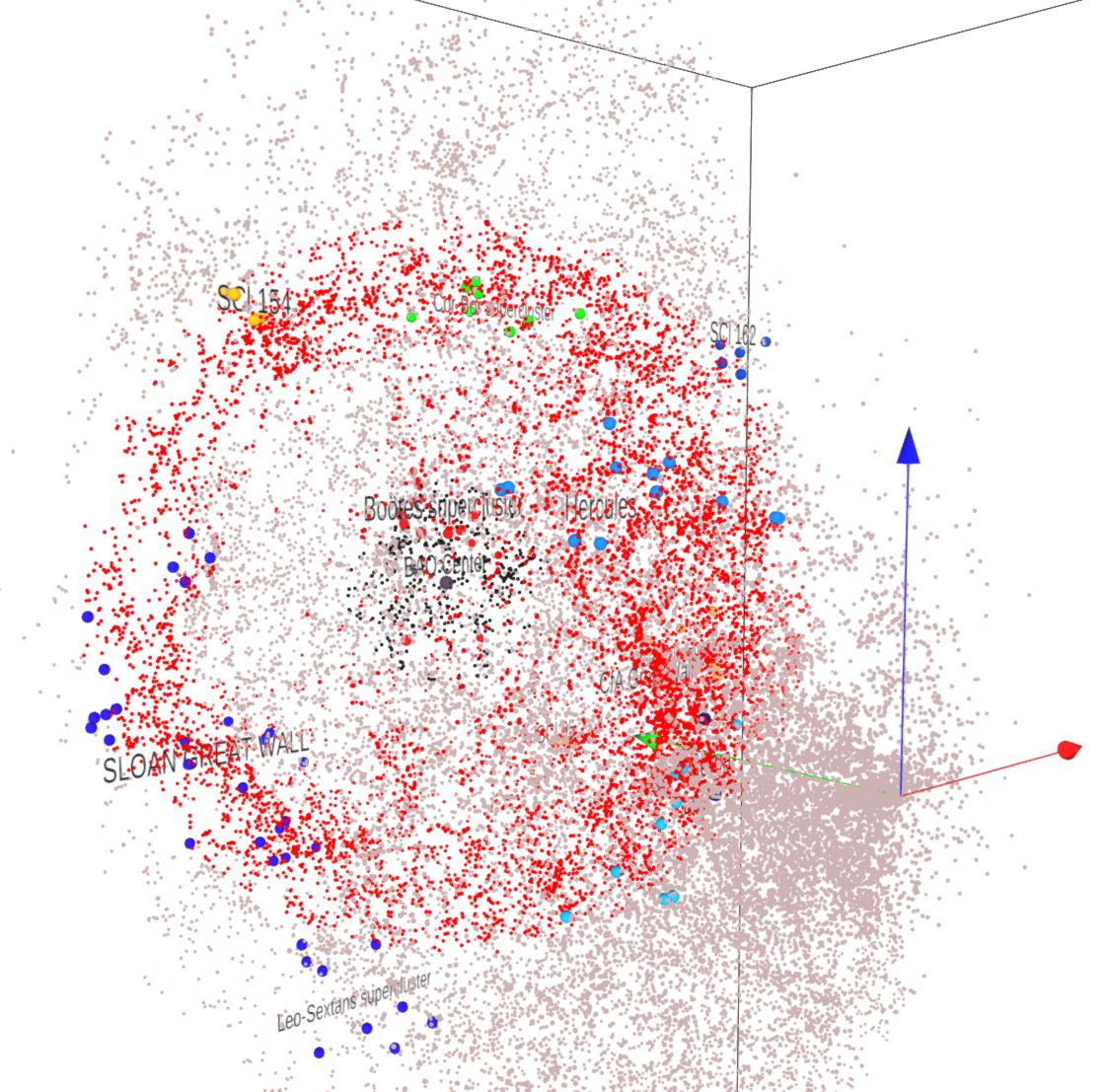}}
\caption{\href{https://sketchfab.com/3d-models/hooleilana-a871418f1a3b4afc89f5cda9f3a70c53}{Interactive 3D visualization of the cosmography of Ho'oleilana}.  All objects in the north galactic hemisphere of the Cosmicflows-4 collection of galaxy groups are seen as points in gray while those lying within the shell of Ho'oleilana, of radius 11,492~\kms\ and width $\pm 2w$ where $w=837$~\kms, are highlighted in red.  Major components in proximity to the shell are highlighted and identified by name.  The Bo\"otes supercluster lies near the center of Ho'oleilana.  Our home location is at the origin of the red, green, blue axes.  These axes have lengths 10,000~\kms\ and are directed toward positive SGX, SGY, SGZ respectively.}
\label{fig:sketchfab}
\end{figure}

\begin{figure}[!]
\centering
\setlength{\fboxsep}{0pt}\fbox{\includegraphics[width=1.0\linewidth]{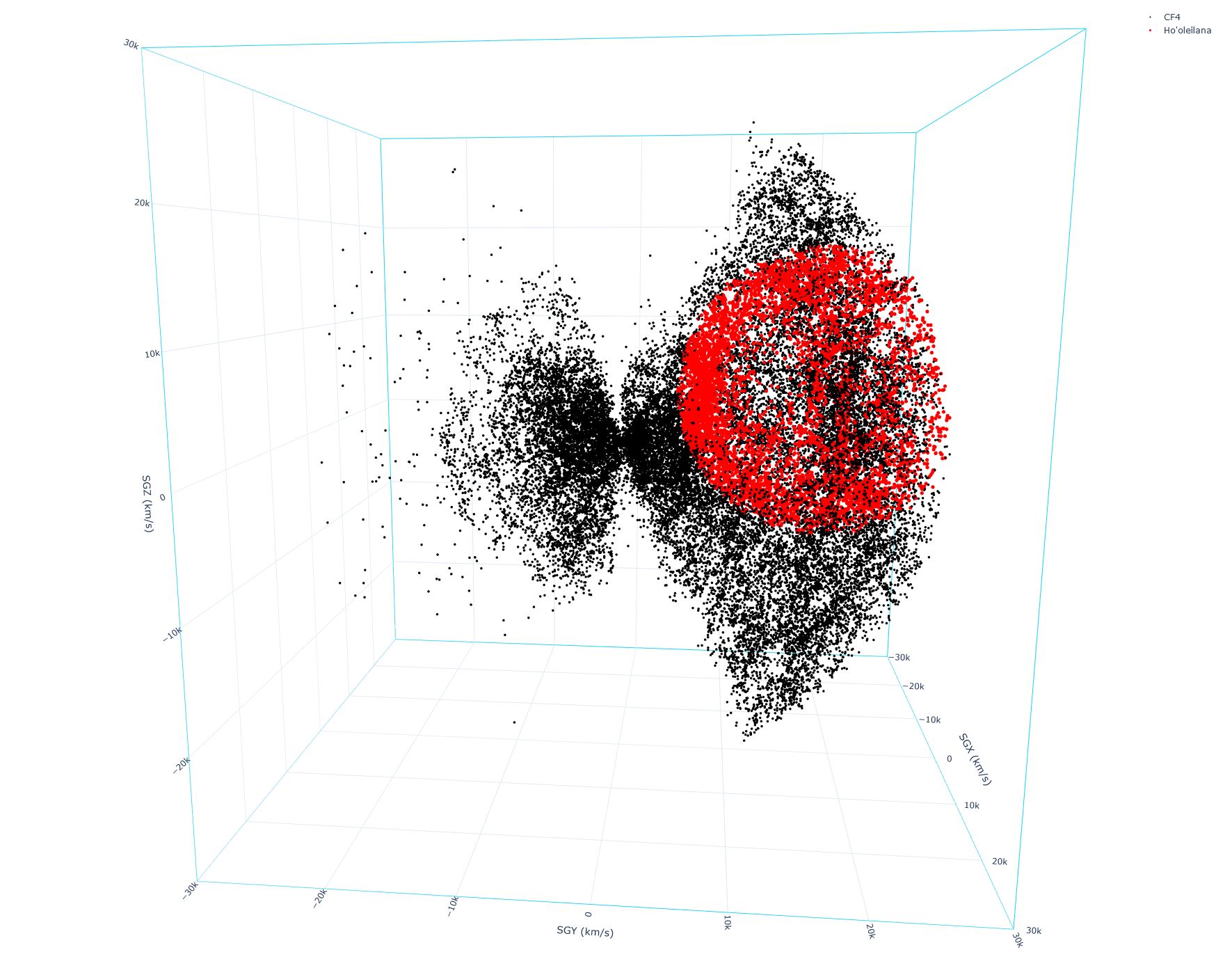}}
\caption{\href{https://irfu.cea.fr/Projets/COAST/Ho'oleilana.html}{Interactive 3D visualization of the Cosmicflows-4 galaxy groups, highlighting those associated with Ho'oleilana}.  Individual galaxy groups are seen as points in black, except those lying within the shell of Ho'oleilana, of radius 11,492~\kms\ and width $\pm 2w$ where $w=837$~\kms, are highlighted in red. Double-click or single-click on the "CF4" or "Ho'oleilana" legends in the upper right corner of the viewer to isolate or hide the corresponding populations. Hover on objects to get positions and PGC "Principal Galaxy Catalog" identifiers.}
\label{fig:membership}
\end{figure}

The extent of Ho'oleilana is revealed in an accompanying video in the animated Figure~\ref{fig:video}. The cosmography of Ho'oleilana is further explored in the interactive Figure~\ref{fig:sketchfab}. The Cosmicflows-4 galaxy groups that lie within the Ho'oleilana shell are colored red in the interactive Figure~\ref{fig:membership}. 
In each of these displays, the galaxy groups are located in 3D by their systemic velocities; ie. in redshift space.


The significance of the detection of Ho'oleilana, its shape, its relation to other previously known structures in the local Universe, and the prominence of the feature compared to the expectations of both a random field of galaxies and simulations with large-scale structure but suppressed BAO, strongly suggest Ho'oleilana is itself a part of the BAO feature rather than a chance alignment. 

Marginalising over the uncertainty in the central position, width and amplitude, we are able to extract a measurement of the ratio of the distance to the center of Ho'oleilana relative to its size predicted by early Universe physics, $D_{v} = 1.63_{-0.08}^{+0.07}\,r_{\mathrm{drag}}$. Fixing to the most likely central location, but using scatter in BAO simulations to infer the error, we find $D_{v} = 1.66\pm0.26\, r_{\mathrm{drag}}$ Given the low redshift of the feature, and adopting a value of the BAO radius $r_{\mathrm{drag}}$ expected from a \citet{Planck+20} cosmological model, these distances can be almost directly converted to constraints on the Hubble Constant, $H_{0} = 76.9^{+8.2}_{-4.8}$ \kmsMpc\ and $H_{0} = 74.7^{+12.4}_{-9.7}$ \kmsMpc\ respectively. These values are more consistent with what is found from other direct local Universe probes --- $H_{0}=69.8\pm0.8\pm1.7$~\kmsMpc\ \citep{Freedman+20}; $H_{0}=73.1\pm1.0$~\kmsMpc\ \citep{Riess+22}; $74.6\pm3.0$~\kmsMpc\ \citep{Tully+23} --- rather than the value of $H_0=67.4\pm0.5$~\kmsMpc\ inferred from propagating the early Universe constraints \citep{Planck+20}. By implication, if Ho'oleilana is representative of the statistical population of BAO, additional late-time physics may be required to increase the expansion rate of the Universe towards the present day. Future deeper data, such as that from the Dark Energy Spectroscopic Instrument \citep{DESI2023} or the 4MOST Hemisphere Survey \citep{4HS2023} may allow for further validation of Ho'oleilana, or the detection of similar structures elsewhere in the nearby Universe. 

\bigskip\noindent
Acknowledgements

We give thanks to collaborators in the assembly of the Cosmicflows-4 catalog of galaxy distances and velocities, with special thanks to H\'el\`ene Courtois, Ehsan Kourkchi, and Khaled Said.  
A perceptive referee challenged us to justify our quantitative uncertainties.
This paper was completed while RBT attended and benefited from discussions with Nick Kaiser\footnote{Sadly, Nick Kaiser passed away on June 13, 2023.} and others at the workshop {\it The Cosmic Web: Connecting Galaxies to Cosmology at High and Low Redshift} at the Kavli Institute of Theoretical Physics, University of California, Santa Barbara.
Funding for the Cosmicflows project has been provided by the US National Science Foundation grant AST09-08846, the National Aeronautics and Space Administration grant NNX12AE70G, and multiple awards to support observations with the Hubble Space Telecope through the Space Telescope Science Institute. CH acknowledges support from the Australian Government through the Australian Research Council’s Laureate Fellowship and Discovery Project funding schemes (projects FL180100168 and DP220101395).

\clearpage

\bibliography{bib}
\bibliographystyle{aasjournal}

\end{document}